\documentclass[conference]{IEEEtran}
\IEEEoverridecommandlockouts
\usepackage{cite}
\usepackage{amsmath,amssymb,amsfonts}
\usepackage{xcolor}
\usepackage{algorithmic}
\usepackage{graphicx}
\usepackage{textcomp}
\usepackage{xcolor}
\def\BibTeX{{\rm B\kern-.05em{\sc i\kern-.025em b}\kern-.08em
    T\kern-.1667em\lower.7ex\hbox{E}\kern-.125emX}}
\begin{document}

\title{The Influence of Cognitive Biases on Architectural Technical Debt
}

\author{
\IEEEauthorblockN{Klara Borowa, Andrzej Zalewski, Szymon Kijas}
\IEEEauthorblockA{\textit{Warsaw University of Technology} \\
\textit{Institute of Control and Computation Engineering}\\
Warsaw, Poland \\
klara.borowa@pw.edu.pl, a.zalewski@ia.pw.edu.pl, szymon.kijas@pw.edu.pl}

}

\maketitle

\begin{abstract}
Cognitive biases exert a significant influence on human thinking and decision-making. In order to identify how they influence the occurrence of architectural technical debt, a series of semi-structured interviews with software architects was performed. The results show which classes of architectural technical debt originate from cognitive biases, and reveal the antecedents of technical debt items (classes) through biases. This way, we analysed how and when cognitive biases lead to the creation of technical debt. We also identified a set of debiasing techniques that can be used in order to prevent the negative influence of cognitive biases. The observations of the role of organisational culture in the avoidance of inadvertent technical debt throw a new light on that issue. 
\end{abstract}


\section{Introduction}
Technical debt is a metaphor first introduced by Cunningham \cite{Cunningham1992} in order to explain the need of refactoring to non-technical stakeholders. It represents experiences that are common to many contemporary software system developers. Namely, the need to compromise between software quality (esp. internal) and other non-technical requirements, such as time-to-release/market. Despite the intense research undertaken to date, the mechanics of the process by which software technical debt arises is still far from being fully explained. This hinders the development and application of systematic technical debt management approaches.  

The main factors that produce the above research challenge are:
\begin{itemize}
    \item the substantial variety of types of technical debt (testing \cite{tom2013exploration}, source code \cite{Amanatidis2018}, architectural \cite{Martini2015}, etc.);
    \item the variety of factors that contribute to the creation of technical debt \cite{Verdecchia2020} \cite{Amanatidis2018};
    \item the social and psychological nature of the phenomenon of technical debt \cite{Brenner2019} and
    \item the intrinsic complexity of the mechanisms underlying the creation of technical debt, resulting from its nature.
\end{itemize}      

Technical debt, whether taken on deliberately or inadvertently, is always rooted in human thinking and/or its limitations. Cognitive biases are an important factor that shape human thinking and decision-making \cite{Mohanani2018}, often distorting its results, making decisions diverge from those fully rational ones. As a result, it is important and necessary to analyse the influences of cognitive bias on the emergence of technical debt, as this aids an understanding of how technical debt arises, and leads to the development of efficient management strategies.

This paper presents our research on the influence of cognitive biases on the occurrence of architectural technical debt (ATD).  It was focused on the following research questions:
\begin{itemize}
    \item \textbf{RQ1.} Do cognitive biases influence the occurrence of architectural technical debt?
    \item \textbf{RQ2.} Which cognitive biases have an impact on architectural technical debt?
    \item \textbf{RQ3.} Which architectural technical debt items are most frequently affected by cognitive biases?
    \item \textbf{RQ4.} What are the antecedents of a harmful influence of cognitive biases on architectural technical debt?
    \item \textbf{RQ5.} What debiasing techniques can be used to minimise the negative effects of cognitive biases?
\end{itemize} 

In order to answer the above questions, we performed semi-structured interviews with 12 architect-practitioners and analysed their outcomes. Detailed information on the research methods can be found in Section \ref{reserch_method_section}, which is preceded by an overview of the current state of research in Section \ref{related_work}. The research outcomes are presented in Section \ref{results_section} and discussed in Section \ref{discussion_section}. The threats to validity are discussed in Section \ref{threats_to_validity_section} and the outcomes summary and further research outlook is presented in Section \ref{conclusion_section}.

\section{Related Work}
\label{related_work}
Cognitive biases, originally observed by Kahneman and Tversky \cite{tversky1974judgment}, \cite{kahneman2011thinking}, are a phenomenon inherent to the human mind. They are rooted  in the duality of the human reasoning process -- according to Kahneman and Tversky -- there are two systems responsible problem resolution that exist and operate within human mind. System 1, which is responsible for quick intuitive decisions based on a limited scope of information. System 2, which is suited for logical and fully rational reasoning on the basis on a broader set of information. System 2 is invoked consciously whenever we analyse and rationally resolve problems. If we do not consciously and carefully consider our decisions (i.e. by employing rational thinking of System 2), System 1 will draw premature conclusions that will not get corrected by System 2. In such case, we can say that a cognitive bias has influenced our reasoning and its outcomes.

The possible influence of cognitive biases on Software Engineering has been already a topic of interest for researchers for over two decades \cite{stacy1995cognitive}. Cognitive biases may possibly influence any Software Engineering activity \cite{Mohanani2018}, be it requirements engineering \cite{zalewski2020cognitive}, design \cite{mohanani2014requirements}, development \cite{Chattopadhyay2020} or testing  \cite{calikli2010empirical}.
Architecture decision-making \cite{Bosch2004} is also not excluded from the impact of cognitive biases \cite{Tang2011}, \cite{VanVliet2016}, \cite{Zalewski2017}. 

The core thesis of this paper, whereby cognitive biases, by distorting the decision-making process, may contribute to taking on technical debt, is at an early stage of research. Only papers  \cite{Chattopadhyay2020}, \cite{Brenner2019} \cite{Verdecchia2020} indicate that such an influence is possible, though there have been no detailed investigations on this topic.

Technical debt has been a topic of extensive study in the recent years. It has resulted in a huge number of papers, including systematic literature reviews \cite{Besker2018c}, \cite{Alfayez2020}, \cite{Becker2018} and even a tertiary study \cite{Rios2018} summarising the state of research on technical debt.

Architectural technical debt (ATD) is the type of TD that occurs as a result of sub-optimal architectural decisions \cite{martini2015danger}. ATD can be especially dangerous since it may hinder the development of future software features \cite{martini2014architecture}.  According to Ernst et al. \cite{Ernst2015} -- "architectural issues are the greatest source of technical debt." The purpose of this paper is to expand the existing knowledge of how architectural technical debt arises by analysing how cognitive biases contribute to its emergence.

\section{Research Method}
\label{reserch_method_section}
\subsection{Cognitive biases}
The number of cognitive biases that possibly can have an influence on software development is vast \cite{Mohanani2018}. As it is not feasible to analyse every cognitive bias in a single study,  we decided to focus on the biases listed as the most relevant for architecture decision-making in the exploratory work of Zalewski et al. \cite{Zalewski2017}. In this work, the researchers attempted to elicit which cognitive biases have the most significant impact on architecture decision-making. These are:
\begin{itemize}
    \item \textit{The framing effect} -- the tendency to judge information and make decisions based on the how the data is presented \cite{Tversky1981}. 
    \item \textit{Confirmation bias} -- this effect influences individuals that have a strong belief they do not want disconfirmed. As such, they search only for information confirming this belief, and ignore any proof that they may be in the wrong \cite{nickerson1998confirmation}. 
    \item \textit{Anchoring bias} -- This bias occurs when one's judgement is strongly influenced by the first piece of information given to them. \cite{Chapman1996} Thus, it often results in individuals having an irrational preference for the first solution/idea that they came up with or heard about from someone else.
    \item \textit{Curse of knowledge bias} -- this cognitive bias manifests itself in experts that consider part of their knowledge as obvious, which then results in miscommunication when they interact with other people \cite{Kennedy1995}.
    \item \textit{IKEA effect bias} -- is the irrational preference for solutions that have been at least partially developed (or assembled) by ourselves \cite{Norton2012}. 
    \item \textit{Parkinson's Law of triviality bias}-- when a disproportionately large amount of time and effort is put into performing trivial tasks and solving trivial problems \cite{northcote1961parkinson}.
    \item \textit{Pro-innovation bias} -- the assumption that innovation is a value in itself. Which means that new solutions should always be adopted everywhere, as soon as possible \cite{Rogers2019}. 
    \item \textit{Planning fallacy bias} -- the tendency to underestimate the time necessary to complete a given task \cite{Pezzo2006}. 
    \item \textit{Bandwagon effect bias} -- the desire to "join the crowd" and do what others do \cite{Leibenstein1950}. This means that popularity becomes the main factor taken into account when choosing between options.
    \item \textit{Irrational escalation bias} -- the irrational impulse to continue wasting resources on an investment that is not cost-effective \cite{Staw2010}. 
    \item \textit{Law of the instrument bias} -- sometimes referred to as the "law of the hammer" since when you own a hammer, everything seems to be a nail \cite{Maslow1966}. This law states that we tend to overuse tools and solutions that we already own or are familiar with.
    \item \textit{Optimism bias} -- the unjustified belief that in our case, in the same scenario, we are more likely to obtain a positive outcome than others\cite{OSulliivan2015}. This effect makes individuals more liable to make risky decisions, despite evidence that it may not be reasonable.
\end{itemize}

\subsection{Architectural Debt items}
Having prepared a list of biases worth exploring, we also had to specify the kinds of architectural technical debt to be researched. There are different approaches to categorising ATD \cite{Besker2018}, \cite{Verdecchia2018}, but for this purpose we decided to use the architectural technical debt items defined by Verdecchia et al. \cite{Verdecchia2020} as the most commonly occurring. We hoped that, since this categorisation emerged from gathering data during interviews, it will also be easily understood by our participants. Those items \cite{Verdecchia2020} include:
\begin{itemize}

\item \textit{Re-inventing the Wheel} -- which manifests itself when we use a self-developed component rather than a stable, verified one that is easily available. 

\item \textit{New Context, Old Architecture} -- which occurs when not enough effort is put into keeping the evolution of the architecture appropriate for its context. 

\item \textit{The Minimum Viable Product (MVP) that stuck} -- which appears when software that was hurriedly developed for a simple temporary solution ends up becoming part of larger system that is still evolving. Architectural gaps of the MVP solution are inherited by the system.

\item \textit{The Workaround that stayed} -- which appears when a temporary workaround is used in order to sidestep architectural constraints. However, it  becomes deeply ingrained in the system and is never removed.

\item \textit{Architectural Lock-in} -- which occurs when a component is so deeply embedded into the system that replacing it would be extremely expensive or even unworkable. 

\item \textit{Source Code architectural technical deb}t -- a type of ATD that has its source solely in the implementation of the solution. 

\end{itemize}

\subsection{Research procedure}
The research method assumed in this paper follows the guidelines for case studies in software engineering by Runeson et al. \cite{Runeson2012}. In order to investigate the influence of cognitive biases on the occurrence of architectural technical debt, we decided to carry out an empirical enquiry based on a set of semi-structured interviews with software architecting practitioners.

The general outline of the interview process that we developed for the purpose of this study, and employed during each of the twelve interviews with architects, was as follows:
\begin{enumerate}
    \item The interviewer asked the participant for their consent to record the interview and to use the acquired data for research purposes.
    \item The researcher obtained statistical data about the participant (age, gender, years of experience, position, company size/domain).
    \item The interviewer introduced the participant to the topic of technical debt and our research.
    \item The participant was provided with the definition of each architectural debt item. Then they were asked if they had ever encountered this type of technical debt and, if so, whether they could describe their experience with it. This is the part of the interview in which the participants had the freedom to provide any information that they wanted and believed to be relevant. 
    \item The interviewer asked the subject if they had any other experiences with technical debt that they had not mentioned yet.
\end{enumerate}

We did not suggest, either before or during the interviews, that cognitive biases may influence technical debt. The participants were informed in advance that we were researching reasons for the occurrence of technical debt, but we could not disclose which reasons we were researching, in order to avoid influencing their answers. After the interviews, if the participant was interested, we disclosed information about our research on cognitive biases. In some cases, this resulted in obtaining additional insights, which were written down for further analysis. 

Almost all of the interviews were conducted in Polish, with the exception of No. 7, conducted in English.

\subsection{Study participants} 
% \begin{table}[htbp]
\begin{table*}[!htp]
\caption{Participant data}
\begin{center}
\begin{tabular}{|c|c|c|c|c|c|c|}
\textbf{No.} & \textbf{Age} & \textbf{Gender} & \textbf{Experience (years)} & \textbf{Position} & \textbf{Company size (employees)} & \textbf{Company domain} \\
\hline
1 & 29 & M & 5 & Software Developer &  over 10 000 &  Electronics\\
2 & 31 & M & 10 & Architect &  around  2 000 &  E-commerce\\
3 & 54 & M & 35 & Chief Operating Officer &  around 1 500 &  High tech\\
4 & 37 & M & 13 & Executive consultant &  around 50 &  Systems integrator\\
5 & 39 & M & 17 & Head of Architects &  around 350 &  Finance\\
6 & 49 & M & 26 & Architect &  around 350 &  Finance\\
7 & 37 & M & 16 & Consultant &  over 10 000 &  Enterprise Software\\
8 & 45 & M & 21 & Chief of Architects &  around 250 &  Systems integrator\\
9 & 36 & M & 15 & Founder and Chief Technology Officer &  around 35 &  Software\\
10 & 37 & F & 15 & Architect &  around 5 000 &  Telecom\\
11 & 40 & M & 15 & Senior Solution Architect &  over 10 000 &  Enterprise Software\\
12 & 37 & M & 12 & Team Leader &  over 10 000 &  Electronics\\
\hline
\end{tabular}
\label{table_participants}
\end{center}
\end{table*}
\label{subsection_participants}

% \begin{table}[htbp]
\begin{table*}[htb!]
\caption{Qualitative analysis codes}
\begin{center}
\begin{tabular}{|p{40mm}|p{30mm}|p{100mm}|}
\textbf{Code category} & \textbf{Code} & \textbf{Definition}  \\
\hline
Cognitive Bias & CB: [bias name] & Occurrence of one of the cognitive biases from the list in Section \ref{reserch_method_section} \\
Architectural technical debt occurrence & ATD: [item type] & Occurrence of technical debt, which can be classified into one of the architectural technical debt items mentioned in Section \ref{reserch_method_section}. This code was to be used only in cases when the participant gave a real-life example of a technical debt occurrence.\\
Architectural technical debt occurrence & ATD: Other & Unclassified occurrence of architectural technical debt\\
Architectural technical debt occurrence influenced by a cognitive bias & CB influencing ATD: [note] & Cases when a cognitive bias directly resulted in the creation of technical debt. The note should contain details of which bias influenced what kind of technical debt items and how.\\
Cognitive bias influencing factor & CB antecedent: [note] & Antecedents of the appearance of cognitive biases. Note should contain a further description.\\
Debasing methods & Debiasing: [note] & Information about interventions that were suggested or performed by the participants, which could result in a debasing effect.\\
\hline
\end{tabular}
\label{table_coding}
\end{center}
\end{table*}

In order to find architects-practitioners, we created an advertisement which we propagated using our private networks. Most of the participants currently work as architects, though some also had prior architecting positions and now worked as leaders/managers/company owners. We also interviewed one software developer, which provided us with a valuable distinct point of view. 
The overall data about the participants is summarised in Table \ref{table_participants}.

\subsection{Analysis Procedure} 
Having obtained the raw data from the interviews, we performed the analysis in the following steps.
\begin{enumerate}
    \item The recorded interviews were transcribed.
    \item We created a coding scheme for analysing the data, using the guidelines of Runeson et al. \cite{Runeson2012}. This codes are presented in Table \ref{table_coding}.
    \item Each of the authors encoded the transcripts independently.
    \item Using the negotiated agreement \cite{Garrison2006} approach, we discussed the coding and incrementally corrected it until we reached unanimity. 
    \item The following metrics were extracted from the transcripts: the number of cases of cognitive bias mentioned by the participants, occurrences of architectural debt, cases of cognitive biases influencing architectural technical debt. Those are presented in Tables \ref{table_ATD_results}, \ref{table_CB_results} and \ref{table_influences_results} in Section \ref{results_section}.
    \item The factors influencing cognitive bias as well as the debiasing methods mentioned by participants were extracted. They are presented in Section \ref{results_section}.
    \item The notes from the interviewer were analysed, in search of any additional data that should be taken into consideration while drawing the conclusions.
    \item The results were discussed and conclusions drawn in a discussion between the authors.
\end{enumerate}

\section{Results}
\label{results_section}

\subsection{Architectural debt items influenced by cognitive biases}
The participants provided us with accounts about their previous projects, in which various architectural debt items could often be observed simultaneously. While analysing the interview transcripts and interviewer notes, we identified 70 specific occurrences of architectural technical debt items, the exact number of occurrences of each ATD item is shown in Table \ref{table_ATD_results}.

% \begin{table}[htbp]
\begin{table}[htb!]
\caption{Technical debt occurrences mentioned by participants}
\begin{center}
\begin{tabular}{|c|c|}
\textbf{Architectural technical debt item} & \textbf{Appearances} \\
\hline
New Context, Old Architecture & 17 \\
Source Code ATD & 13 \\
The Workaround that stayed & 12 \\
Architectural Lock-in & 10\\
Re-inventing the Wheel & 8 \\
The Minimum Viable Product that stuck & 6 \\
Other (4 different types of ATD) & 4 \\
\hline
\end{tabular}
\label{table_ATD_results}
\end{center}
\end{table}
Despite providing the participants with the definitions of architectural technical debt and the specific architectural debt items, they often mentioned situations that were not cases of architectural technical debt, or which fit the definition of a different technical debt item. A common mistake was the false belief that the use of an old technology is synonymous with technical debt, which does not have to be the case. 

The most commonly occurring ATD item was "New Context, Old Architecture". This specific kind of technical debt appears naturally, by itself, over time, if not enough effort is given to periodically update, upgrade, change or refactor the architecture. This classical problem has already been described in one of Lehman's laws of software evolution \cite{Lehman1980}, which says that at some point of time, perpetually evolving systems reach a threshold when it is no longer cost-effective to evolve further without carrying out a major system's reconstruction. Many of our participants observed, that this moment often passes unnoticed. 

We found several instances of technical debt that did not exactly fit any of the categories of ATD items specified by Verdecchia et al. \cite{Verdecchia2020}. These technical debt items are:
\begin{itemize}
    \item \textit{Choosing an obsolete solution that should not be used in the current circumstances at the start of the project}. Participant No 5 explained that in his company, decision-makers only consider aged solutions as "safe enough" to be used.
    \item \textit{Reusing a component in a setting, in which it does not fit the given problem}. Participant No 1 told us how one of his colleagues focused on using readily-made solutions so much, that it resulted in choosing a solution completely unsuitable for their problem.
    \item \textit{Using a proven architectural solution in a new context, in which it is not suitable}. Participant No 4 explained a situation in which they had to deal with data on the client's customers, for which they used a readily-made component that integrated all the aspects of every customer's data. The problem appeared when it turned out that not all the customers wanted to have all of their data connected to a single account, since they may want to create many accounts for various purposes.
    \item \textit{Transfer of organisational debt onto architectural technical debt}. Organisational debt occurs when key decisions (such as writing down contracts, defining strategies or assigning responsibilities) are not made in time.  This, in turn, may affect key design decisions, which have to be made with incomplete data about the problem at hand. Participant No. 3 gave us an example of a situation when a state-owned system had to be deployed before certain key political decisions were made. This resulted in the need to redo the basic components of the system.
\end{itemize}
We did not observe a correlation between the participants’ experience, position or their company's domain and the number of ATD items they observed. An interesting case of this came from participants Nos 5 and 6. Both worked in the same organisation, No 5 had nine years of experience less than No 6, but participant No 5 gave us nine examples of ATD item occurrences, while participant No 6 mentioned only two. \\

\subsection{Cognitive biases that influence ATD items}
By analysing the transcripts and interviewer notes, we found 155 occurrences of cognitive biases. The exact numbers for each bias are shown in Table \ref{table_CB_results}. It was not unusual for many biases to influence a single ATD item, which often occurs consecutively in a cascade of irrational decisions, such as:
\begin{itemize}
    \item A decision-maker heard that a specific technology is popular, which led him to believe that it may be useful in his case (Bandwagon effect)
    \item He met with a salesman of this specific solution, who only informed him about the beneficial aspects of the solution, which persuaded him to buy it (Framing effect)
    \item Despite the disadvantages of this solution, it was used simply because it had already been paid for (Irrational escalation)
    \item Which led to an "Architectural Lock-in" because this component was so specific and deeply embedded in the system that it turned out extremely difficult to replace.
\end{itemize}
In our analysis we took into account not only the biases of the architects and developers, but also those that influenced other stakeholders involved in the development and maintenance process – such as the management and the clients. 

Most of the biases mentioned as possibly crucial in the work of Zalewski et al. \cite{Zalewski2017} had a notable influence on ATD. An exception here is Parkinson's law of triviality. This specific bias may impact the time spent on certain tasks, but it does not seem to significantly change the final outcomes, and so has little magnitude when it comes to causing ATD. 
% \begin{table}[htbp]
\begin{table}[htb!]
\caption{Cognitive biases present in the participants' accounts}
\begin{center}
\begin{tabular}{|c|c|}
\textbf{Cognitive bias} & \textbf{Appearances} \\
\hline
Anchoring &	24                           \\
Bandwagon effect 	& 8                  \\
Confirmation bias	& 19                 \\
Curse of knowledge	& 14                 \\
IKEA effect	& 14                         \\
Irrational escalation	& 11             \\
Law of the instrument	& 10             \\
Optimism bias	& 20                     \\
Parkinson's Law of triviality	& 2      \\
Planning fallacy	& 10                 \\
Pro-innovation bias	& 13                 \\
The framing effect	& 10				 \\
\hline
\end{tabular}
\label{table_CB_results}
\end{center}
\end{table}

\subsection{Influence of cognitive biases on ATD items} 
In this Section we discuss in-depth how particular ATD items were impacted by cognitive biases. Table \ref{table_influences_results} presents the exact number of times that a certain cognitive bias influenced particular ATD items. The data contained in this table does not add up to the data from Table \ref{table_CB_results} and Table \ref{table_ATD_results}, because a specific bias occurrence may have influenced a single ATD item more than once, and one ATD item may have been influenced by more than one cognitive bias. \\
If the influence of a particular bias on a specific ATD item was reported at least three times, we explored thoroughly the relationship between that bias and the ATD item. 

% \begin{table}[htbp]
\begin{table*}[!htp]
\caption{Cognitive biases influencing ATD items}
\begin{center}
\begin{tabular}{|c|p{17mm}|p{17mm}|p{17mm}|p{17mm}|p{17mm}|p{17mm}|p{17mm}|}
\textbf{Cognitive Bias} & \textbf{New Context, Old Architecture} & \textbf{Source Code ATD} & \textbf{The Workaround that stayed} & \textbf{Architectural Lock-in} & \textbf{Re-inventing the Wheel} & \textbf{Minimum Viable Product that stuck} & \textbf{Other} \\
\hline
Anchoring	&	7   & 5 & 4 & 6 & 4 & 1 & 0                   \\
Bandwagon effect	&	0   & 1 & 1 & 1 & 1 & 0 & 0                    \\
Confirmation bias	&	2   & 2 & 5 & 4 & 5 & 1 & 1             \\
Curse of knowledge	&	2   & 2 & 2 & 4 & 2 & 0 & 0              \\
IKEA effect	&	3   & 3 & 1 & 2 & 3 & 1 & 0              \\
Irrational escalation	&	7   & 1 & 2 & 0 & 1 & 1 & 0                \\
Law of the instrument	&	1   & 3 & 2 & 3 & 0 & 0 & 0 \\
Optimism bias	&	3 & 3 & 3 & 5 & 2 & 4 & 1 \\
Parkinson's Law of triviality &	0 & 1 & 2 & 0 & 0 & 0 & 0 \\
Planning fallacy	&	3 & 4 & 4 & 3 & 1 & 1 & 1\\
Pro-innovation bias	&	1 & 1 & 2 & 4 & 4 & 2 & 1 \\
The framing effect	&	1 & 2 & 2 & 3 & 1 & 1 & 0 \\
\hline
\end{tabular}
\label{table_influences_results}
\end{center}
\end{table*}

\begin{enumerate}
    \item \textit{New Context, Old Architecture} \\
The significant impact of cognitive biases on architectural technical debt can be clearly observed when researching the causes of this ADT item. 

Four participants experienced a situation when a solution was chosen simply because it was the first possible one that came to notice (\textit{anchoring}), and even though it was not cost-effective and did not enable the further evolution of the product, resources were persistently being wasted on it (\textit{irrational escalation}). 

Participant No. 1 for example, was involved in a project where an open source solution was chosen to create a simple dashboard for the end user. Unfortunately, as the solution evolved and expanded, the source code of this component had to be forked. Ultimately, the participant's team introduced an enormous amount of changes and became the maintainer of this newly created solution. This increased the team's workload with the maintenance efforts.

Too often, a component was used after the support for it had expired, which either left the component without maintenance or forced the client to take the path of expensive, dedicated individual support. Participant No. 11 told us about a few cases of systems made for the public sector, when his clients needed this kind of individual maintenance. This could have been prevented by properly preparing for the time when this problem was bound to occur, but often no such precautions are taken, and decisions are made without long-term planning.

In general, it seems that the decision to start from scratch with completely new architecture is made reluctantly. This hesitancy can be motivated by many cognitive biases: the \textit{IKEA effect} when the old solution was made by the decision-maker's organisation (participant No. 9 called a product his "precious business baby"), or the \textit{optimism bias} which may make the decision-makers believe that no harm may come to them (for example, when using a system that is not maintained properly).

\item \textit{Source Code ATD} 
Source code ATD is most often influenced by \textit{anchoring}. When solving a problem, the very basic, satisficing approach is widely present. Citing participant No. 11 – "If something stupid works, then it is not stupid." This is especially apparent in organisations in which decisions on how to implement certain components are left entirely to individual developers whose ideas are never challenged. This was the case in an example provided by participant No. 2, who had the displeasure of "inheriting" a completely unscalable solution, used as a basis for a key e-commerce platform developed by his company.

This problem of satisficing decision-making is strengthened by other cognitive biases as well. The \textit{IKEA effect} makes developers choose (or even copy) from their own previous work, the \textit{law of instrument} makes them use only tools that they are familiar with. The \textit{confirmation bias} makes them blind to information that their decisions may be wrong, an effect that is often enhanced by the \textit{optimism bias}.

Participant No. 12 encountered a combination of all of this biases in the form of an enormous Bash based solution, used to deploy changes to the production environment. The author of the solution was comfortable with Bash, and did not consult (nor was he challenged) his solution's design with anyone. He did not take into account that anyone else may ever need to read, understand or change his code. This resulted in the creation of an enormous set of Bash scripts that were extremely hard to comprehend to anyone besides its author.

All these problems become even more relevant when not enough time is given for thorough consideration, which is an intermediate effect of the \textit{planning fallacy} – when the planned time for tasks was too short.

\item \textit{The Workaround that stayed} \\
A substantial number of workarounds generally come from two beliefs. Namely, that there is no choice, and that fast fixes are a normal and proper way of problem solving. Individuals that firmly believe in either of them often do not put any additional effort into considering the, often lacking, rationale behind their "fixes" (\textit{confirmation bias}). \\
Having come up with an idea for a simple workaround, they are satisfied that the problem will be promptly resolved, and do not search for any alternatives (\textit{anchoring}). \\
Participant No. 3 gave us an interesting example, of an organisation that routinely used a complicated set of workarounds while processing accounting data. Only when a new integrated accounting system was introduced,  the organisation did realise that they have been producing faulty financial reports for the last 5 years.
This kind of a mindset is further strengthened when they are not given enough time, because the the need for such such tasks has not been foreseen (\textit{planning fallacy}).\\
However the workaround does solve the immediate problem, so there is no urgent need to change the state of things. This approach, heavily tainted by the \textit{optimism bias}, was displayed by Participant No. 9 with the words "we will hopefully come back to it one of these days."

\item \textit{Architectural Lock-in} \\
The most common pattern behind the occurrence of the Architectural lock-in ATD item was a combination of \textit{anchoring} and \textit{optimism bias}. Firstly, due to anchoring, the first satisficing architectural solution was chosen. Then, even though they did not have any experience with the solution, development teams simply took a leap of faith (\textit{optimism bias}) and used the solution without further consideration of whether it may be difficult to replace later.\\
This would not have become such a serious problem if individuals did not have a tendency to choose risky innovative solutions (\textit{pro-innovation bias}), or if they took time to consider the disadvantages of their architectural concepts (\textit{confirmation bias}).\\
Participant No. 12 provided us with the following example that illustrates this problem. The maintenance team in his organisation needed a ticketing system. They found an simple open source solution on GitHub that looked satisfactory and choose it, blindly believing that it would be the proper one. They did not take into account that the component may require changes in the future and that it was PHP-based (no one in that team had prior experience in PHP). When they needed to expand the solution, they found themselves "locked-in" this particular component, while not having the skills necessary for its further development.

Additionally, we noticed that decision-makers often relied on data from salesmen, which is of course always prepared in a way that shows the offered solutions in a positive light (\textit{framing effect}).  Participant No. 3 particularly stressed how "salesmen should never be trusted".  Even if decision-makers attempted to obtain information from experts within their own organisation – the experts had a tendency to omit key information during meetings, because of the false belief that such knowledge is obvious (\textit{curse of knowledge}). 

\item \textit{Re-inventing the Wheel} \\
The Re-inventing the wheel ATD item was mainly observed by our participants in the context of younger, inexperienced team members, as well as in small companies, especially start-ups.
Lacking prior experience, and with the possibility to start something new from scratch, ambitious young people often fall into the temptation of creating a solution they would have full control over (thus, \textit{anchoring} on that single aspect), something they could call "their own" (\textit{IKEA effect}). They want to be pioneers (\textit{pro-innovation bias}), despite often not possessing and not searching for already existing knowledge. This frequently results in re-inventing the metaphorical wheel. 

Participant No. 7 told us about a case when he worked in a small company, with a colleague that he described as an "IT geek". This coworker developed a web application framework on his own. He frequently applied it when creating products for the company. After he changed job, this undocumented framework was left without its core maintainer.

The unwillingness to face the reality that someone may have already had the same idea and properly put it into effect, and thus the inability to find and use ready-made components, is a symptom of the dangerous influence of the \textit{confirmation bias}. 

\item\textit{The Minimum Viable Product that stuck} \\
We found that the MVP that stuck ATD item was the least likely to appear from the list defined by Verdecchia et al. \cite{Verdecchia2020}. In our participants' accounts, MVPs rarely "got stuck", which means that our observations for this type of ATD were limited. In the case of most MVPs mentioned by our participants, these solutions were either abandoned as prototypes when a superior solution was found/developed, or these MVPs were expanded and matured over time. 

For an MVP remaining as it is over an extended period of time, someone had to make a mistake while estimating a very short lifespan for the product (\textit{optimism bias}). Participant No. 5 told us that this usually happened when small programs were written in a hurry to perform simple tasks like processing/converting text files or uploading/downloading them.

\end{enumerate}

\subsection{Cognitive bias antecedents (RQ4)}

Having identified the biases that influenced the generation of technical debt, we made an in-depth analysis of the participants’ accounts in search of information on why these cognitive biases occurred. Cognitive biases, as a phenomenon inherent to the human mind, cannot be completely avoided. However, certain factors can amplify the influence of cognitive biases on architecting activities and their outcomes. These factors, namely, the antecedents of cognitive biases, can be divided into seven groups:
\begin{enumerate}
\item \textit{Individual's emotional state } \\
We found that certain feelings often precede the appearance of biases. These are:

\begin{enumerate}
    \item Fear of: change, responsibility, consequences and of starting from square one (precedes \textit{anchoring, confirmation bias, irrational escalation});
    \item Shame, especially of one's past mistakes (precedes\textit{anchoring, irrational escalation});
    \item Feeling a lack of agency (makes individuals less likely to challenge the ideas of others and provide a debiasing effect);
    \item Haste (makes individuals more susceptible to\textit{all} biases, especially \textit{the planning fallacy});
\end{enumerate}
Fear and shame have a notably destructive effect. One of our participants attempted to register information about the technical debt in their organisation. This turned out to be difficult because employees, even managers, were unwilling to share information about the technical debt they were responsible for. They were afraid of consequences and ashamed of their previous mistakes, which then hindered the process of actively managing technical debt. Individuals experiencing great fears are unlikely to change their behaviour, which makes them even more susceptible to biases such as \textit{anchoring, confirmation bias and irrational escalation}. \\
Feeling a lack of agency is especially relevant in large organisations, in which individuals often feel that they have no influence on any decisions that are being made. Because of that, they simply remain indifferent, do not challenge others’ decisions, and end up mindlessly following orders.

\item \textit{Individual's personality traits} \\
There were two kinds of personality types that were overly prone to cognitive biases. The extremely ambitious and confident individuals and their opposite – the reserved ones that lacked assertiveness. The overconfident architects tended to make fast decisions without deeper consideration (this makes \textit{all} cognitive biases more likely to appear), while the taciturn team members tended to follow them blindly. In this way, the possibility of exerting a debiasing effect on their colleagues is lost. It also makes them more prone to the \textit{bandwagon effect}.

\item \textit{Individual's mistakes} \\
We observed some common mistakes that foreshadowed the appearance of cognitive biases. Those included:
\begin{itemize}
    \item The basic lack of knowledge or experience required to make decisions in a certain area (if sufficient knowledge is not obtained, \textit{any} cognitive biases are more likely to occur).
    \item Not performing any search for alternative solutions (\textit{anchoring, confirmation bias, IKEA effect, law of the instrument, pro-innovation bias}).
    \item Considering only a limited part of a complex problem, while ignoring the global impact of the solution (\textit{anchoring, confirmation bias, curse of knowledge, optimism bias, planning fallacy}).
    \item Limiting the planning only to the short term (\textit{planning fallacy, optimism bias}).
    \item Habits (\textit{confirmation bias, anchoring, irrational escalation, IKEA effect, law of instrument}).
    \item The optimistic belief that mistakes are only made by others (makes \textit{all} biases more likely to appear).
\end{itemize}

An interesting detail is that even seemingly good habits, such as using proven architectural patterns, may turn out to be harmful. In the case of this particular habit, sometimes a design pattern may end up being used in unsuitable circumstances.

\item \textit{Organisational antecedents} \\
The overall environment in which the project is being developed and maintained has a crucial impact. We observed several factors that impacted the frequency of bias appearance:
\begin{itemize}
    \item Organisational culture: too lax (which strengthens all biases in individual employees) or too harsh (which impacts the biases of management and leaders).
    \item Frequent changes of management staff that impedes long-term planning (influences \textit{all} biases)
    \item Lack of standards and procedures (\textit{all} biases).
    \item Unclear separation of duties, especially when it is not clear who is responsible for which decisions (this means that nobody provides a debiasing effect when decisions are made in this area).
    \item Short-sighted cost/profit optimisation as a default approach – investing only in areas that give immediate profit (\textit{irrational escalation, optimism bias, anchoring}).
    \item Lack of motivation for optimising the developed solutions – especially in the case of short-term cooperation with clients (optimism bias).
    \item Faulty use of agile development practices – empowered by the belief that any problems can be fixed in further iterations (influences \textit{all} biases, since decisions are made in each iteration by all the parties involved in the project)  
\end{itemize}

A valuable observation that we made was that, in most of the interviews, cognitive biases appeared as a consequence of an organisational culture that was either too lax or too harsh. When the culture in the organisation was too casual, which is often the case in startups (participant No 9 provided us with such insights), individuals are often left to make key decisions alone. If these decisions are influenced by cognitive biases, no one challenges them and thus various faulty decisions are made – mainly by young, overambitious team members. This may lead to many biased decisions like choosing trendy solutions (\textit{bandwagon effect}) or using only tools that the decision-maker is familiar and comfortable with (\textit{law of instrument}).

On the other hand, if the organisation's culture is authoritarian, giving little voice to the employees in lower positions in the hierarchy, then possible biases of the higher-ups are never challenged or corrected. In such cases, decision-makers are more susceptible to the enticements of salesmen (\textit{framing effect}), or do not have information that would allow them to plan the time-frames for projects properly (\textit{planning fallacy)}.

\item \textit{Communicational antecedents} \\
Many biases emerge as an after-effect of communication problems. This most commonly happens when specialists from different domains interact, although even close co-workers are not free from this problem. These problems are often fuelled by the \textit{curse of knowledge}, which makes individuals more likely to omit crucial information that could be obtained from others. \\
Constructive criticism and challenging the ideas underlying the decisions of others is not standard in every team. It often means that crucial decisions are never discussed openly. This means that valuable debiasing opportunities are lost, which in turn makes biased decisions more likely to occur.
Additionally, sometimes decisions made during the initial negotiation phase of a project are made without consulting technical specialists, which leads to overly strict deadlines and sometimes absurd contractual arrangements. Participant No 11 told us that, in the case of projects made for the state, they commonly found that the price and time-frame for the project were made absurdly low and short during negotiations, simply in order to gain the customer, in hope that profit could be increased later and fixes could be made during the maintenance phase.

\item \textit{Knowledge vaporisation} \\
For any decision to be rational, it is essential for the decision-makers to have proper knowledge about the issue at hand. However, it is a well-known issue in the area of architectural knowledge management that knowledge is not always documented and tends to vanish with the employees that leave the company. This makes decision-makers more prone to the effect of \textit{all} cognitive biases, since they are frequently forced to make decisions based solely on their instincts.
\item \textit{External} \\
Some antecedents to cognitive biases are completely beyond our control. The most prevalent is the current popularity of certain solutions – the deciding factor behind the \textit{bandwagon effect}. \\
Although, often forgotten, an important source of possible problems also lies in politics and the current legal status. A proper interpretation and understanding of the law is not an easy task and often leads to dangerous misconceptions. One of our participants provided us with an account in which they had to create a system before a particular law came into force. This law had been incorrectly understood by the developers (the \textit{curse of knowledge} influenced their communication with legal specialists), which resulted in the need to perform a set of quick fixes and workarounds shortly after the system became available. \\ Frequent changes in the law, influenced by politics, may also force numerous technically challenging modifications to existing systems. \textit{Every} such decision is susceptible to the influence of cognitive biases – in this case, the politicians' biases.

\end{enumerate}

\subsection{Possible debiasing methods (RQ5)}
During the interviews, our participants spontaneously gave us hints, how to avoid arising certain ATD items.  Additionally, the participants that scarcely encounter specific ATD items, usually mentioned why they believe that this item does not occur often in their environment. All this combined together, in many cases, can be interpreted as a set of possible debiasing techniques.

We gathered this into a set of bias prevention treatments:
\begin{enumerate}
    \item Ensuring double-checking and challenging all decisions and their underlying ideas. Trying to find downsides of any idea as standard – this will make the critique feel less personal and is therefore more effective. 
    \item Developing an environment based on trust, in which employees can voice their opinions and admit to their mistakes – knowing that they will receive help, not scorn. The modern approaches to agile / servant leadership address this issue.
    \item Explicitly gathering information about alternatives before making decisions. Presenting them to others and asking for their evaluation. This is reminiscent of some components of architecture evaluation methods.
    \item Creating procedures and standards to limit low-quality reckless alterations of the solution and enable periodical refactoring/changes of the system to fit the ever-changing business context in which it is used.
    \item Creating documentation and passing on knowledge. At a bare minimum, this does not require much resources, it might involve recording meetings in which information is shared and decisions are made. 
    \item Explicitly registering all accounts of TD in the organisation and making plans for a time when it will be dealt with.
    \item Periodically checking whether any new TD has occurred and whether any old TD needs to be paid – maybe support expired, or all the people with the relevant knowledge have moved on and are no longer present.
    \item Clearly defining and recording who is responsible for which part of the project's scope. This will make the process of obtaining information and looking for help more straightforward. It will also minimise the problem of the individuals in charge avoiding responsibility.
\end{enumerate}

\section{Discussion}
\label{discussion_section}
As the presented results have no direct equivalents (the possible influence of biases on technical debt has been merely mentioned \cite{Brenner2019},\cite{Verdecchia2020}, \cite{Chattopadhyay2020}), it is necessary to relate these results to broader research on technical debt, cognitive biases in SE, antecedents and management techniques for technical debt.

Firstly, we partially confirmed the findings of Zalewski et al. \cite{Zalewski2017}, since almost all the biases (with the exception of Parkinson's law of triviality) that they recognised as notable for architecture decision-making, were found to have had a significant influence on ATD.
Furthermore, the biases that we have identified as most commonly influencing ATD (anchoring, optimism and confirmation bias) have already been identified as having a significant influence on software engineering activities \cite{Mohanani2018}.

In the field of architectural decision-making, it has already been noticed that cognitive biases distort the decision-making process \cite{Manjunath2018} by strengthening the effects of pre-existing problems/mistakes. As such, the antecedents for cognitive biases are bound to be, at least partially, similar to previously discovered causes of ATD. The problems of time pressure, lack of documentation, unsuitable architectural decisions and human factors, as specified by Verdecchia et al. \cite{Verdecchia2020}, are similar to many of the antecedents that we identified. The issue of miscommunication between specialists of different domains, and its influence on technical debt, has also previously been addressed \cite{Stablein2018}. 

The debiasing methods that we propose only affect debt created inadvertently, since debt deliberately taken on is usually a result of a rational management strategy \cite{Besker2018b}. 
The debiasing methods that we proposed give an interesting new perspective to the problem of managing architectural technical debt – they can be taken as a set of instructions on how to manage an organisation that would be less susceptible to ATD. ATD management so far, as indicated by Besker et al. \cite{Besker2018c}, suffers from a lack of proper management guidelines. A set of strategies has only recently been proposed \cite{Verdecchia2020}, \cite{Besker2020}. 

\section{Threats to Validity}
\label{threats_to_validity_section}
As with every study, certain issues might pose a threat to the validity of our findings. Having this in mind, we attempted to minimise the effects of such threats.
Since the research is qualitative, and our goal was to conduct an exploratory in-depth analysis, we only took into account the experiences of our 12 participants. To prevent this from being a problem, we attempted to make this group as varied as possible – we interviewed people that held different positions, had varying levels of experience (from 5 to 35 years), worked in companies of different sizes (from start-ups to large corporations), and whose organisations had diverse domains. 

Still, in order to further confirm the validity of this research, it would be useful to expand it with more participants, and possibly using a different research methodology (like the think aloud protocol method \cite{fonteyn1993description}).

Since our participants were not experts in the field of technical debt, they often presented examples of cases that were not actually occurrences of technical debt. Furthermore, even if their example was indeed a case of technical debt, they confused various ATD items withe each other. To ensure that such mistakes did not have an undue influence on our results, we searched for the ATD items (coded them from the transcriptions) without taking into account the ATD item category that the participant believed their example belonged to. 

Since it may be possible for a single researcher to make a mistake during the coding – for example, to observe a cognitive bias that did not actually occur – the interview transcriptions were analysed and coded by us separately, and then the findings were confronted using the negotiated agreement approach \cite{Garrison2006} 

To prevent our participants from forcibly searching for cognitive biases in their experience, we only asked them to explain the rationale behind the decisions made in their projects. They were informed about our cognitive bias related research only after the interviews.

Finally, cognitive biases often overlap and interact with each other. Which means that their influence on ATD items may not always be straightforward. We did not analyse the dynamics of bias' interaction in this paper. 

\section{Conclusion and Research outlook}
\label{conclusion_section}
Our research achieved the following:
\begin{itemize}
    \item We assessed that cognitive biases definitely have a significant influence on the creation of architectural technical debt (RQ1).
    \item We determined that most significant biases that impact architectural technical debt are anchoring, optimism and confirmation bias. Nevertheless, the influences of the curse of knowledge, the IKEA Effect, irrational escalation, pro-innovation bias, planning fallacy, the framing effect and the bandwagon effect are also noticeable.
    \item  We assessed that cognitive biases affect all of the ATD items indicated by Verdecchia et al. \cite{Verdecchia2020}; nevertheless, the most frequently affected item turned out to be "New Context, Old Architecture" (RQ3) and the least influenced one appeared to be "the MVP that stuck".
    \item The most common antecedents of cognitive biases that influence ATD have been identified (RQ4).
    \item A number of debiasing techniques have been proposed (RQ5).
\end{itemize}

Our research revealed also that the organisation's culture is often an important factor that influences the creation of technical debt, since most of the antecedents and the discovered debiasing methods are connected with how the organisation is managed, and the frame of mind of the organisation's members. 

In order to minimise the amount of unwanted architectural technical debt, organisations should remove the fear of admitting software deficiencies and introduce trust into the company's culture. As noticed by Besker et al. \cite{Besker2020}, the penalising approach to managing architectural technical debt is the least effective one. 

An atmosphere of thrust and camaraderie would enable individuals to provide a debiasing effect to each other. The ideal environment would be one in which challenging the ideas of others is commonplace, in the form of sensible, non-judgemental critique. If meaningful critique of each other’s ideas becomes commonplace, employees are less likely to feel threatened by it and to actually start making use of each other’s suggestions.

In an organisation founded on trust, there should be space to admitting one's mistakes. When a problem is detected, everyone should focus on solving it together, instead of looking for scapegoats.

Further research could include:
\begin{itemize}
    \item further confirming our findings with proper quantitative data;
    \item how team / organisational culture influences the emergence of inadvertent technical debt;
    \item in-depth research on particular antecedents' and biases' influence on ATD;

\end{itemize}

\bibliographystyle{myIEEEtran.bst}
\bibliography{refs}

\begin{thebibliography}{10}
\providecommand{\url}[1]{#1}
\csname url@samestyle\endcsname
\providecommand{\newblock}{\relax}
\providecommand{\bibinfo}[2]{#2}
\providecommand{\BIBentrySTDinterwordspacing}{\spaceskip=0pt\relax}
\providecommand{\BIBentryALTinterwordstretchfactor}{4}
\providecommand{\BIBentryALTinterwordspacing}{\spaceskip=\fontdimen2\font plus
\BIBentryALTinterwordstretchfactor\fontdimen3\font minus
  \fontdimen4\font\relax}
\providecommand{\BIBforeignlanguage}[2]{{%
\expandafter\ifx\csname l@#1\endcsname\relax
\typeout{** WARNING: IEEEtran.bst: No hyphenation pattern has been}%
\typeout{** loaded for the language `#1'. Using the pattern for}%
\typeout{** the default language instead.}%
\else
\language=\csname l@#1\endcsname
\fi
#2}}
\providecommand{\BIBdecl}{\relax}
\BIBdecl

\bibitem{Cunningham1992}
W.~Cunningham, ``{The WyCash portfolio management system},'' in
  \emph{Proceedings of the Conference on Object-Oriented Programming Systems,
  Languages, and Applications, OOPSLA}, vol. Part F1296, 1992. doi:
  10.1145/157709.157715. ISBN 0897916107 pp. 29--30.

\bibitem{tom2013exploration}
E.~Tom, A.~Aurum, and R.~Vidgen, ``An exploration of technical debt,''
  \emph{Journal of Systems and Software}, vol.~86, no.~6, pp. 1498--1516, 2013.

\bibitem{Amanatidis2018}
\BIBentryALTinterwordspacing
T.~Amanatidis, N.~Mittas, A.~Chatzigeorgiou, A.~Ampatzoglou, and L.~Angelis,
  ``{The developer's dilemma: Factors affecting the decision to repay code
  debt},'' in \emph{2018 IEEE/ACM International Conference on Technical Debt
  (TechDebt)}, vol.~5, 2018. doi: 10.1145/3194164.3194174. ISBN 9781450357135.
  ISSN 02705257 pp. 62--66. [Online]. Available:
  \url{https://doi.org/10.1145/3194164.3194174}
\BIBentrySTDinterwordspacing

\bibitem{Martini2015}
\BIBentryALTinterwordspacing
A.~Martini and J.~Bosch, ``{The Danger of Architectural Technical Debt:
  Contagious Debt and Vicious Circles},'' in \emph{Proceedings - 12th Working
  IEEE/IFIP Conference on Software Architecture, WICSA 2015}, 2015. doi:
  10.1109/WICSA.2015.31. ISBN 9781479919222 pp. 1--10. [Online]. Available:
  \url{https://www.researchgate.net/publication/273769557}
\BIBentrySTDinterwordspacing

\bibitem{Verdecchia2020}
R.~Verdecchia, P.~Kruchten, and P.~Lago, ``{Architectural Technical Debt : A
  Grounded Theory},'' \emph{European Conference on Software Architecture
  (ECSA)}, 2020.

\bibitem{Brenner2019}
R.~Brenner, ``{Balancing resources and load: Eleven nontechnical phenomena that
  contribute to formation or persistence of technical debt},'' in
  \emph{Proceedings - 2019 IEEE/ACM International Conference on Technical Debt,
  TechDebt 2019}, 2019. doi: 10.1109/TechDebt.2019.00013. ISBN 9781728133713
  pp. 38--47.

\bibitem{Mohanani2018}
R.~Mohanani, I.~Salman, B.~Turhan, P.~Rodriguez, and P.~Ralph, ``{Cognitive
  Biases in Software Engineering: A Systematic Mapping Study},'' \emph{IEEE
  Transactions on Software Engineering}, vol. 5589, no.~c, 2018. doi:
  10.1109/TSE.2018.2877759

\bibitem{tversky1974judgment}
A.~Tversky and D.~Kahneman, ``Judgment under uncertainty: Heuristics and
  biases,'' \emph{science}, vol. 185, no. 4157, pp. 1124--1131, 1974.

\bibitem{kahneman2011thinking}
D.~Kahneman, \emph{Thinking, fast and slow}.\hskip 1em plus 0.5em minus
  0.4em\relax Macmillan, 2011.

\bibitem{stacy1995cognitive}
W.~Stacy and J.~MacMillan, ``Cognitive bias in software engineering,''
  \emph{Communications of the ACM}, vol.~38, no.~6, pp. 57--63, 1995.

\bibitem{zalewski2020cognitive}
A.~Zalewski, K.~Borowa, and D.~Kowalski, ``On cognitive biases in requirements
  elicitation,'' in \emph{Integrating Research and Practice in Software
  Engineering}.\hskip 1em plus 0.5em minus 0.4em\relax Springer, 2020, pp.
  111--123.

\bibitem{mohanani2014requirements}
R.~Mohanani, P.~Ralph, and B.~Shreeve, ``Requirements fixation,'' in
  \emph{Proceedings of the 36th International Conference on Software
  Engineering}, 2014, pp. 895--906.

\bibitem{Chattopadhyay2020}
\BIBentryALTinterwordspacing
S.~Chattopadhyay, N.~Nelson, A.~Au, N.~Morales, C.~Sanchez, R.~Pandita, and
  A.~Sarma, ``{A Tale from the Trenches : Cognitive Biases and Software
  Development},'' in \emph{International Conference on Software Engineering
  (ICSE)}, 2020. doi: 10.1145/3377811.3380330. ISBN 9781450371216 pp. 654--665.
  [Online]. Available: \url{https://doi.org/10.1145/3377811.3380330}
\BIBentrySTDinterwordspacing

\bibitem{calikli2010empirical}
G.~Calikli and A.~Bener, ``Empirical analyses of the factors affecting
  confirmation bias and the effects of confirmation bias on software
  developer/tester performance,'' in \emph{Proceedings of the 6th International
  Conference on Predictive Models in Software Engineering}, 2010, pp. 1--11.

\bibitem{Bosch2004}
J.~Bosch, ``{Software architecture: The next step},'' \emph{Lecture Notes in
  Computer Science (including subseries Lecture Notes in Artificial
  Intelligence and Lecture Notes in Bioinformatics)}, vol. 3047, pp. 194--199,
  2004. doi: 10.1007/978-3-540-24769-214

\bibitem{Tang2011}
A.~Tang, ``{Software designers, are you biased?}'' \emph{Proceedings -
  International Conference on Software Engineering}, no. January 2011, pp.
  1--8, 2011. doi: 10.1145/1988676.1988678

\bibitem{VanVliet2016}
H.~van Vliet and A.~Tang, ``{Decision making in software architecture},''
  \emph{Journal of Systems and Software}, vol. 117, pp. 638--644, 2016. doi:
  10.1016/j.jss.2016.01.017

\bibitem{Zalewski2017}
A.~Zalewski, K.~Borowa, and A.~Ratkowski, ``{On cognitive biases in
  architecture decision making},'' in \emph{Lecture Notes in Computer Science
  (including subseries Lecture Notes in Artificial Intelligence and Lecture
  Notes in Bioinformatics)}, vol. 10475 LNCS, 2017. doi:
  10.1007/978-3-319-65831-59. ISBN 9783319658308. ISSN 16113349 pp. 123--137.

\bibitem{Besker2018c}
\BIBentryALTinterwordspacing
T.~Besker, A.~Martini, and J.~Bosch, ``{Managing architectural technical debt:
  A unified model and systematic literature review},'' \emph{Journal of Systems
  and Software}, vol. 135, pp. 1--16, 2018. doi: 10.1016/j.jss.2017.09.025.
  [Online]. Available: \url{https://doi.org/10.1016/j.jss.2017.09.025}
\BIBentrySTDinterwordspacing

\bibitem{Alfayez2020}
\BIBentryALTinterwordspacing
R.~Alfayez, W.~Alwehaibi, R.~Winn, E.~Venson, and B.~Boehm, ``{A systematic
  literature review of technical debt prioritization},'' in \emph{Proceedings
  of the 3rd International Conference on Technical Debt}, vol.~10.\hskip 1em
  plus 0.5em minus 0.4em\relax ACM, 2020. doi: 10.1145/3387906.3388630. ISBN
  9781450379601 pp. 1--10. [Online]. Available:
  \url{https://doi.org/10.1145/3387906.3388630}
\BIBentrySTDinterwordspacing

\bibitem{Becker2018}
\BIBentryALTinterwordspacing
C.~Becker, R.~Chitchyan, S.~Betz, and C.~McCord, ``{Trade-off decisions across
  time in technical debt management: A systematic literature review},'' in
  \emph{2018 IEEE/ACM International Conference on Technical Debt
  (TechDebt)}.\hskip 1em plus 0.5em minus 0.4em\relax ACM, 2018. doi:
  10.1145/3194164.3194171. ISBN 9781450357135. ISSN 02705257 pp. 85--94.
  [Online]. Available: \url{https://doi.org/10.1145/3194164.3194171}
\BIBentrySTDinterwordspacing

\bibitem{Rios2018}
N.~Rios, M.~Mendonça, and R.~Spínola, ``A tertiary study on technical debt:
  Types, management strategies, research trends, and base information for
  practitioners,'' \emph{Information and Software Technology}, vol. 102, 06
  2018. doi: 10.1016/j.infsof.2018.05.010

\bibitem{martini2015danger}
A.~Martini and J.~Bosch, ``The danger of architectural technical debt:
  Contagious debt and vicious circles,'' in \emph{2015 12th Working IEEE/IFIP
  Conference on Software Architecture}.\hskip 1em plus 0.5em minus 0.4em\relax
  IEEE, 2015, pp. 1--10.

\bibitem{martini2014architecture}
A.~Martini, J.~Bosch, and M.~Chaudron, ``Architecture technical debt:
  Understanding causes and a qualitative model,'' in \emph{2014 40th EUROMICRO
  Conference on Software Engineering and Advanced Applications}.\hskip 1em plus
  0.5em minus 0.4em\relax IEEE, 2014, pp. 85--92.

\bibitem{Ernst2015}
\BIBentryALTinterwordspacing
N.~A. Ernst, S.~Bellomo, I.~Ozkaya, R.~L. Nord, and I.~Gorton, ``{Measure it?
  Manage it? Ignore it? Software practitioners and technical debt},'' in
  \emph{2015 10th Joint Meeting of the European Software Engineering Conference
  and the ACM SIGSOFT Symposium on the Foundations of Software Engineering,
  ESEC/FSE 2015 - Proceedings}, 2015. doi: 10.1145/2786805.2786848. ISBN
  9781450336758 pp. 50--60. [Online]. Available:
  \url{http://github.com/neilernst/td-survey}
\BIBentrySTDinterwordspacing

\bibitem{Tversky1981}
A.~Tversky and D.~Kahneman, ``{The framing of decisions and the psychology of
  choice},'' \emph{Science}, 1981. doi: 10.1126/science.7455683

\bibitem{nickerson1998confirmation}
R.~S. Nickerson, ``Confirmation bias: A ubiquitous phenomenon in many guises,''
  \emph{Review of general psychology}, vol.~2, no.~2, pp. 175--220, 1998.

\bibitem{Chapman1996}
G.~B. Chapman and B.~H. Bornstein, ``{The more you ask for, the more you get:
  Anchoring in personal injury verdicts},'' \emph{Applied Cognitive
  Psychology}, 1996. doi:
  10.1002/(SICI)1099-0720(199612)10:6<519::AID-ACP417>3.0.CO;2-5

\bibitem{Kennedy1995}
J.~Kennedy, ``{Debiasing in the Audit Curse of Knowledge Judgment},'' \emph{The
  Accounting Review}, 1995.

\bibitem{Norton2012}
M.~I. Norton, D.~Mochon, and D.~Ariely, ``{The IKEA effect: When labor leads to
  love},'' \emph{Journal of Consumer Psychology}, 2012. doi:
  10.1016/j.jcps.2011.08.002

\bibitem{northcote1961parkinson}
C.~Northcote~Parkinson, ``Parkinson's law: or the pursuit of progress,'' 1961.

\bibitem{Rogers2019}
E.~M. Rogers, A.~Singhal, and M.~M. Quinlan, ``{Diffusion of innovations},'' in
  \emph{An Integrated Approach to Communication Theory and Research, Third
  Edition}, 2019. ISBN 9781351358712

\bibitem{Pezzo2006}
M.~V. Pezzo, J.~A. Litman, and S.~P. Pezzo, ``{On the distinction between
  yuppies and hippies: Individual differences in prediction biases for planning
  future tasks},'' \emph{Personality and Individual Differences}, 2006. doi:
  10.1016/j.paid.2006.03.029

\bibitem{Leibenstein1950}
H.~Leibenstein, ``{Bandwagon, snob, and veblen effects in the theory of
  consumers' demand},'' \emph{Quarterly Journal of Economics}, 1950. doi:
  10.2307/1882692

\bibitem{Staw2010}
B.~M. Staw, ``{The escalation of commitment: An update and appraisal},'' in
  \emph{Organizational Decision Making}, 2010.

\bibitem{Maslow1966}
A.~H. Maslow, \emph{{The psychology of science; a reconnaissance}}, 1966. ISBN
  National Library: 0354146 LCCN: 66-11479

\bibitem{OSulliivan2015}
O.~P. O'Sulliivan, ``{The Neural Basis of Always Looking on the Bright Side},''
  \emph{Dialogues in Philosophy, Mental and Neuro}, 2015.

\bibitem{Besker2018}
\BIBentryALTinterwordspacing
T.~Besker, A.~Martini, and J.~Bosch, ``{Technical debt cripples software
  developer productivity: A longitudinal study on developers' daily software
  development work},'' in \emph{2018 IEEE/ACM International Conference on
  Technical Debt (TechDebt)}, vol.~10, 2018. doi: 10.1145/3194164.3194178. ISBN
  9781450357135. ISSN 02705257 pp. 105--114. [Online]. Available:
  \url{https://doi.org/10.1145/3194164.3194178}
\BIBentrySTDinterwordspacing

\bibitem{Verdecchia2018}
R.~Verdecchia, ``{Architectural Technical Debt Identification: Moving
  Forward},'' \emph{Proceedings - 2018 IEEE 15th International Conference on
  Software Architecture Companion, ICSA-C 2018}, pp. 43--44, 2018. doi:
  10.1109/ICSA-C.2018.00018

\bibitem{Runeson2012}
\BIBentryALTinterwordspacing
P.~Runeson, M.~H{\"{o}}st, A.~Rainer, and B.~Regnell, \emph{{Case Study
  Research in Software Engineering: Guidelines and Examples}}, 2012. ISBN
  9781118104354. [Online]. Available: \url{www.wiley.com.}
\BIBentrySTDinterwordspacing

\bibitem{Garrison2006}
D.~R. Garrison, M.~Cleveland-Innes, M.~Koole, and J.~Kappelman, ``{Revisiting
  methodological issues in transcript analysis: Negotiated coding and
  reliability},'' \emph{Internet and Higher Education}, vol.~9, no.~1, pp.
  1--8, 2006. doi: 10.1016/j.iheduc.2005.11.001

\bibitem{Lehman1980}
M.~M. Lehman, ``{Programs, Life Cycles, and Laws of Software Evolution},''
  \emph{Proceedings of the IEEE}, vol.~68, no.~9, pp. 1060--1076, 1980. doi:
  10.1109/PROC.1980.11805

\bibitem{Manjunath2018}
A.~Manjunath, M.~Bhat, K.~Shumaiev, A.~Biesdorf, and F.~Matthes, ``{Decision
  Making and Cognitive Biases in Designing Software Architectures},''
  \emph{Proceedings - 2018 IEEE 15th International Conference on Software
  Architecture Companion, ICSA-C 2018}, pp. 52--55, 2018. doi:
  10.1109/ICSA-C.2018.00022

\bibitem{Stablein2018}
\BIBentryALTinterwordspacing
T.~Stablein, D.~Berndt, and M.~Mullarkey, ``{Technical debt-related information
  asymmetry between finance and IT},'' in \emph{2018 IEEE/ACM International
  Conference on Technical Debt (TechDebt)}, 2018. doi: 10.1145/3194164.3194180.
  ISBN 9781450357135. ISSN 02705257 pp. 134--137. [Online]. Available:
  \url{https://doi.org/10.1145/3194164.3194180}
\BIBentrySTDinterwordspacing

\bibitem{Besker2018b}
T.~Besker, A.~Martini, R.~{Edirisooriya Lokuge}, K.~Blincoe, and J.~Bosch,
  ``{Embracing technical debt, from a startup company perspective},''
  \emph{Proceedings - 2018 IEEE International Conference on Software
  Maintenance and Evolution, ICSME 2018}, pp. 415--425, 2018. doi:
  10.1109/ICSME.2018.00051

\bibitem{Besker2020}
\BIBentryALTinterwordspacing
T.~Besker, A.~Martini, and J.~Bosch, ``{Carrot and stick approaches when
  managing technical debt},'' in \emph{Proceedings of the 3rd International
  Conference on Technical Debt}, 2020. doi: 10.1145/3387906.3388619. ISBN
  9781450379601 pp. 21--30. [Online]. Available:
  \url{https://doi.org/10.1145/3387906.3388619}
\BIBentrySTDinterwordspacing

\bibitem{fonteyn1993description}
M.~E. Fonteyn, B.~Kuipers, and S.~J. Grobe, ``A description of think aloud
  method and protocol analysis,'' \emph{Qualitative health research}, vol.~3,
  no.~4, pp. 430--441, 1993.

\end{thebibliography}

\end{document}